\documentclass[aps,prl,twocolumn,showpacs,nofootinbib,amsmath,amssymb,amsfonts,superscriptaddress]{revtex4-1}

\usepackage{graphicx}
\usepackage{epsfig,latexsym}

\usepackage[usenames]{color}
\RequirePackage{xspace} \allowdisplaybreaks

\usepackage{natbib}
\usepackage{url}
\usepackage{lineno}

\usepackage{bm}
\usepackage{dcolumn}
\usepackage{graphicx}
\usepackage{graphics}
\usepackage[latin1]{inputenc}
\usepackage{latexsym}
\usepackage{rotating}
\usepackage{xspace} 

\usepackage{ulem}
\usepackage{outlines}
\usepackage{enumitem}
\setenumerate[1]{label=\Roman*.}
\setenumerate[2]{label=\Alph*.}
\setenumerate[3]{label=\roman*.}
\setenumerate[4]{label=\alph*.}

\definecolor {darkgreen}{rgb}{0.2,0.7,0.2}

\begin{document}

\newcommand{\nl}{\nonumber\\}

\newcommand{\ans}{ansatz }
\newcommand{\mat}[4]{\left(\begin{array}{cc}{#1}&{#2}\\{#3}&{#4}
\end{array}\right)}
\newcommand{\matr}[9]{\left(\begin{array}{ccc}{#1}&{#2}&{#3}\\
{#4}&{#5}&{#6}\\{#7}&{#8}&{#9}\end{array}\right)}
\newcommand{\matrr}[6]{\left(\begin{array}{cc}{#1}&{#2}\\
{#3}&{#4}\\{#5}&{#6}\end{array}\right)}
\newcommand{\cvb}[3]{#1^{#2}_{#3}}
\def\lsim{\raise0.3ex\hbox{$\;<$\kern-0.75em\raise-1.1ex
e\hbox{$\sim\;$}}}
\def\gsim{\raise0.3ex\hbox{$\;>$\kern-0.75em\raise-1.1ex
\hbox{$\sim\;$}}}
\def\abs#1{\left| #1\right|}
\def\simlt{\mathrel{\lower2.5pt\vbox{\lineskip=0pt\baselineskip=0pt
           \hbox{$<$}\hbox{$\sim$}}}}
\def\simgt{\mathrel{\lower2.5pt\vbox{\lineskip=0pt\baselineskip=0pt
           \hbox{$>$}\hbox{$\sim$}}}}
\def\unity{{\hbox{1\kern-.8mm l}}}
\newcommand{\eps}{\varepsilon}
\def\ep{\epsilon}
\def\ga{\gamma}
\def\Ga{\Gamma}
\def\om{\omega}
\def\omp{{\omega^\prime}}
\def\Om{\Omega}
\def\la{\lambda}
\def\La{\Lambda}
\def\al{\alpha}
\newcommand{\ov}{\overline}
\renewcommand{\to}{\rightarrow}
\renewcommand{\vec}[1]{\mathbf{#1}}
\newcommand{\vect}[1]{\mbox{\boldmath$#1$}}
\def\tm{{\widetilde{m}}}
\def\mcirc{{\stackrel{o}{m}}}
\newcommand{\Dm}{\Delta m}
\newcommand{\dm}{\varepsilon}
\newcommand{\tanb}{\tan\beta}
\newcommand{\nbar}{\tilde{n}}
\newcommand\PM[1]{\begin{pmatrix}#1\end{pmatrix}}
\newcommand{\up}{\uparrow}
\newcommand{\down}{\downarrow}
\def\omE{\omega_{\rm Ter}}
%

\newcommand{\Dsusy}{{susy \hspace{-9.4pt} \slash}\;}
\newcommand{\DCP}{{CP \hspace{-7.4pt} \slash}\;}
\newcommand{\mc}{\mathcal}
\newcommand{\gr}{\mathbf}
\renewcommand{\to}{\rightarrow}
\newcommand{\gtc}{\mathfrak}
\newcommand{\wh}{\widehat}
\newcommand{\br}{\langle}
\newcommand{\kt}{\rangle}

\newcommand{\Pl}{{\mbox{\tiny Pl}}}
\newcommand{\stat}{{\mbox{\tiny stat}}}
\newcommand{\tot}{{\mbox{\tiny tot}}}
\newcommand{\sys}{{\mbox{\tiny sys}}}
\newcommand{\GW}{{\mbox{\tiny GW}}}
\newcommand{\ny}[1]{\textcolor{blue}{\it{\textbf{ny: #1}}} }
\newcommand{\am}[1]{\textcolor{red}{\it{\textbf{am: #1}}} }

\newcommand{\Hor}{{\mbox{\tiny H}}}
\newcommand{\BH}{{\mbox{\tiny BH}}}
\newcommand{\HL}{{\mbox{\tiny HL}}}
\newcommand{\Bondi}{{\mbox{\tiny Bondi}}}
\newcommand{\DM}{{\mbox{\tiny DM}}}
\newcommand{\Rel}{{\mbox{\tiny Rel}}}
\newcommand{\IGM}{{\mbox{\tiny IGM}}}
\newcommand{\ISM}{{\mbox{\tiny ISM}}}
\newcommand{\CMB}{{\mbox{\tiny CMB}}}
\newcommand{\DE}{{\mbox{\tiny DE}}}
\newcommand{\tidal}{{\mbox{\tiny tidal}}}
\newcommand{\nonspin}{{\mbox{\tiny no-spin}}}
\newcommand{\spinalign}{{\mbox{\tiny spin-aligned}}}
\newcommand{\comp}{{\mbox{\tiny comp}}}


\def\lsim{\mathrel{\mathop  {\hbox{\lower0.5ex\hbox{$\sim$}
\kern-0.8em\lower-0.7ex\hbox{$<$}}}}}
\def\gsim{\mathrel{\mathop  {\hbox{\lower0.5ex\hbox{$\sim$}
\kern-0.8em\lower-0.7ex\hbox{$>$}}}}}

\def\nn{\\  \nonumber}
\def\de{\partial}
\def\brf{{\mathbf f}}
\def\bbf{\bar{\bf f}}
\def\bF{{\bf F}}
\def\bbF{\bar{\bf F}}
\def\bA{{\mathbf A}}
\def\bB{{\mathbf B}}
\def\bG{{\mathbf G}}
\def\bI{{\mathbf I}}
\def\bM{{\mathbf M}}
\def\bY{{\mathbf Y}}
\def\bX{{\mathbf X}}
\def\bS{{\mathbf S}}
\def\bb{{\mathbf b}}
\def\bh{{\mathbf h}}
\def\bg{{\mathbf g}}
\def\bla{{\mathbf \la}}
\def\bmu{\mathbf m }
\def\by{{\mathbf y}}
\def\bmu{\mbox{\boldmath $\mu$} }
\def\bsig{\mbox{\boldmath $\sigma$} }
\def\bunity{{\mathbf 1}}
\def\cA{{\cal A}}
\def\cB{{\cal B}}
\def\cC{{\cal C}}
\def\cD{{\cal D}}
\def\cF{{\cal F}}
\def\cG{{\cal G}}
\def\cH{{\cal H}}
\def\cI{{\cal I}}
\def\cL{{\cal L}}
\def\cN{{\cal N}}
\def\cM{{\cal M}}
\def\cO{{\cal O}}
\def\cR{{\cal R}}
\def\cS{{\cal S}}
\def\cT{{\cal T}}
\def\eV{{\rm eV}}
%

\title{Gravitational Instability of Exotic Compact Objects}

\author{Andrea Addazi}
\author{Antonino Marcian\`o}
\affiliation{Center for Field Theory and Particle Physics \& Department of Physics, Fudan University, 200433 Shanghai, China}
\author{Nicol\'as Yunes}
\affiliation{eXtreme Gravity Institute, Department of Physics, Montana State University, Bozeman, MT 59717, USA}

\begin{abstract}
\noindent

Exotic compact objects with physical surfaces a Planckian distance away from where the horizon would have been are inspired in quantum gravity. Most of these objects are defined by a classical spacetime metric, such as boson stars, gravastars and wormholes. We show that these classical objects are gravitationally unstable because accretion of ordinary and dark matter, and gravitational waves forces them to collapse to a black hole by the Hoop conjecture. To avoid collapse, their surface must be a macroscopic distance away from the horizon or they must violate the null energy condition.

\end{abstract}

\maketitle

\noindent 

\vspace{0.3cm}
\noindent
{\emph{Introduction}}.~The recent discovery of gravitational waves~\cite{Abbott:2016blz,Abbott:2016nmj,Abbott:2017oio,Abbott:2017vtc,Abbott:2017gyy} has triggered new interesting ideas in black holes physics. These waves have the potential to probe fundamental physics beyond General Relativity~\cite{Yunes:2013dva,Yunes:2016jcc,TheLIGOScientific:2016src,Abbott:2018lct,LIGOScientific:2019fpa}, such as the speed and the dispersion relation of these waves~\cite{Mirshekari:2011yq}, gravitational parity invariance~\cite{Alexander:2007kv,Yunes:2010yf,Alexander:2017jmt}, and Lorentz invariance in gravity~\cite{Hansen:2014ewa}. Within such exciting possibilities, one is but forced to wonder whether hints of quantum gravity could also be observed or their existence constrained with gravitational wave observations. 

One such hint is the possibility that massive compact objects may seem like black holes but may not actually be black holes because they may lack an event horizon. Such black hole mimickers are called Exotic Compact Objects (ECOs)~\cite{Cardoso:2019rvt}, and their lack of an event horizon implies the existence of a physical surface, though the latter may be diffuse as in boson stars~\cite{Ruffini:1969qy}. ECOs are inspired from quantum gravity ideas and they may be classified into classical and quantum ECOs. Classical ECOs are defined explicitly in terms of a spacetime metric and an equation of state, as in the case of gravastars~\cite{Mazur:2001fv}, while quantum ECOs are still heuristic, as in the case of fuzzballs~\cite{Mathur:2005zp}. The features of quantum ECOs depend on the specific quantum gravity framework used to describe them, and thus, the issues encountered in quantizing gravity percolate into ambiguities in the quantum ECO's metric representation and its equation of state. Therefore, we here study classical ECOs only (and refer to them as simply  ECOs for conciseness), which are also the most studied ones in the gravitational wave literature,  and not give their quantum counterparts any further consideration. 

Although precisely how ECOs form and how generic their formation is remains unclear, they have been a playground for theorists to explore how gravitational waves could be used to distinguish between the coalescence of black holes and that of other exotica and to detect signatures of quantum gravity at horizon scales. One such signatures concerns tidal Love numbers, which characterize the degree of deformation of an object in the presence of an external tidal field~\cite{Cardoso:2017cfl,Maselli:2017cmm}. When compact objects are in a coalescing binary system, their mutual tidal deformations are encoded in the gravitational waves emitted. Therefore, since black holes have zero tidal Love numbers, any non-zero measurement could serve as smoking gun evidence for an ECO, provided it can be distinguished from a neutron star~\cite{Sennett:2017etc}, even if this measurement cannot be used to then infer the location of the surface to Planckian precision~\cite{Addazi:2018uhd}. 

Another signature concerns echoes~\cite{Cardoso:2016oxy,Cardoso:2017njb,Cardoso:2017cqb}. Let us assume that two ECOs in a binary system coalesce and form another, deformed ECO that settles down to equilibrium through the emission of gravitational waves. The replacement of the event horizon with a surface changes the exterior spacetime geometry, allowing some of the gravitational waves emitted after merger to bounce back from a hump in the effective potential, and the rest to tunnel through and escape to spatial infinity. Some of the now-incoming gravitational waves can be absorbed by the ECO, while the remaining amount will bounce off its surface and head back out to spatial infinity. The process can then be repeated, resulting in a series of echoes of the ringdown gravitational waves~\cite{Cardoso:2016oxy,Cardoso:2017njb,Cardoso:2017cqb}.   
 
\vspace{0.3cm}
\noindent
{\emph{The stability of an ECO.}} How realistic are ECOs and how seriously should one take such alternatives? This is a difficult question to answer because there are no field equations whose solution leads to these objects. One can of course concoct a carefully designed spacetime, and then insert it into the Einstein equations to compute the stress-energy tensor that leads to such a metric. But at least in General Relativity, such stress-energy tensor will violate the energy conditions. Without field equations one can also not answer whether the generic collapse of matter leads to the formation of such objects. But at the very least, one must demand an unconditional requirement: stability.  

Ultracompact objects that lack an event horizon typically suffer from instabilities. One of them is associated with the existence of a stable light ring~\cite{Cunha:2017qtt}, which implies that electromagnetic and gravitational perturbations decay at most logarithmically, forcing them to ``pile up'' near the light ring and increasing the energy densities until a trapped surface forms~\cite{Keir:2014oka,Cardoso:2014sna}. The other instability is associated with the existence of an ergosphere~\cite{Cardoso:2007az}, because negative energy states that exist inside it can be trapped in the effective potential and cascade to even more negative states. Both instabilities can be circumvented if one allows the ECO surface to be partially absorbing, so that the timescale for the onset of the instability becomes very long~\cite{Maggio:2018ivz}.

We here focus on the stability of ECOs against gravitational collapse due to interactions with their medium, and show that they generically encounter insurmountable obstacles that prevent them from having a surface too close to their would-be horizons. This, in turn, places a theoretical maximum on their gravitational compactness that limits how similar to black holes they can be. If this limit is broken, we show that ECOs collapse to BHs during, or even before, they enter the inspiral phase of coalescence. These obstacles render ECOs, and their subsequent echoes, theoretically unfavored.     
 
The obstacles described above rely on Thorne's Hoop conjecture~\cite{ThorneMagic,Misner:1974qy}, which states that a compact object collapses to form a black hole when it fits inside a certain critical 2-sphere, a surface of revolution constructed from rotating a circular hoop of a certain critical circumference $C = 2 \pi R_{\Hor}$, where $R_{\Hor}$ is the radius of the would-be horizon. The conjecture has been verified in various scenarios, including non-spherical system~\cite{Nakamura:1988zq,Malec:1991nf}, colliding black holes~\cite{Ida:1998qt}, non-time-symmetric initial data~\cite{Yoshino:2001ik}, colliding pp-waves~\cite{Yoshino:2007yb}, charged curved spacetimes~\cite{Hod:2018cxr}, and even in loop quantum gravity~\cite{Anza:2017dkd}. The conjecture has also been more rigorously reformulated through new definitions of the circumference~\cite{Chiba:1994ir}, trapped circles~\cite{Senovilla:2007dw}, Birkhoff's invariant~\cite{Gibbons:2009xm}, and the Brown-York mass~\cite{Murchadha:2009lav}.

Let us then assume that ECOs are immersed in a medium that forces them to gain mass, while maintaining their radius approximately constant, as is the case for neutron stars. If the accretion rate $\dot{M}$ is sufficiently weak and constant, the amount of mass gained in an amount of time $T$ is simply $M-M_{0} = \delta M = \dot{M} T$. If the accretion rate is not constant, then $\delta M$ must be calculated through integration, as we show later. Regardless of whether the rate is constant or not, if enough mass is accreted, the ECO collapses to a BH by the Hoop conjecture, so prevent this, one must have
\begin{align}
\label{eq:Hoop}
R = R_{\Hor}(M_{0}) + \delta R > R_{\Hor}(M_{0}+\delta M)
\end{align}
where $R$ is the ECO radius, $\delta R$ is the distance of the ECO surface from the horizon radius, which can be cast as an invariant measure of length following \cite{Addazi:2018uhd}, and the right-hand side of the inequality comes about due to the Hoop conjecture and must be evaluated at the new ECO mass of $M = M_{0} + \delta M$.  Assuming $\delta M \ll M_{0}$ is small and linearizing 
\begin{align} 
\label{eq:min-radius}
\delta R &> \frac{G}{c^{2}} \delta M
\left[\frac{1 - \chi^{2} + (1 - \chi^{2})^{1/2}}{(1-\chi^{2})^{1/2}}\right]\,,
\end{align}
where we have assumed that the dimensionless spin parameter $\chi = |\vec{S}/M_{0}^{2}|$, with $\vec{S}$ its spin angular momentum, remains roughly constant during this process. The above places a limit on the maximum compactness that ECOs can have
\begin{align}
\label{eq:max-comp}
C &:= \frac{M_{0}}{R} \approx C_{\BH} \left(1 - \frac{\delta R}{R_{\Hor}(M_{0})}\right)\,,
\end{align}
where $C_{\BH} := M_{0}/R_{\Hor}(M_{0})$. In what follows, we consider different mechanisms that can increase the mass of an ECO, and then we derive the minimum distance that the ECO surface can be from its would-be horizon. 

\vspace{0.3cm}
\noindent
{\emph{Accretion of matter by an ECO in a non-empty Universe.}} ECOs, if they exist, are not alone. Like everything else in the Universe, ECOs should be surrounded by dark energy, dark matter, an intergalactic medium and an interstellar medium. Let us then calculate how much mass-energy ECOs accrete from these media.

Given that ECOs would be completely immersed in these media, we can model their accretion through the Bondi-Hoyle framework. We assume then that the accretion  rate is given by (see e.g.~\cite{PPR,BH,Shi,BHL})
\begin{equation}
\label{eq:bondi-rate}
\dot{M}_{\Bondi}= \frac{4\pi G^2 M_{0}^2 \rho}{(c_s^{2} + {v}_\infty^2 )^{\frac{3}{2}} } \,,
\end{equation}
where $M_{0}$ is the mass of the ECO prior to accretion, $v_{\infty}$ is the asymptotic velocity of the infalling fluid, $\rho$ is the density of the medium, and $c_{s}$ is the speed of sound in the fluid. This expression interpolates between the regimes in which the infalling matter is subsonic and supersonic (outside and inside the Bondi radius), reducing to the Hoyle-Lyttleton accretion rate when $c_{s} \ll v_{\infty}$. 

Let us now apply this accretion rate to matter in the interstellar, intergalactic and intracluster medium. The interstellar medium consists of gas in various forms that fills the space between stars in a galaxy. The density is $(10^{-4},10^{-2}) \times 10^{-21} \,  {\rm kg}/{\rm{m^{3}}}$ and the sound speed is about $10^{-3} c$ at a temperature of $10^{6}$ K, with $c$ the speed of light~\cite{Ferriere:2001rg}. Using this and the Bondi-Hoyle rate for a supermassive ECO with mass $M_{0} = 10^{6} M_{\odot}$, we then find
\begin{equation}
\label{eq:bondi-est-ISM}
\dot{M}_{\Bondi}^{\ISM}= \left(10^{-9},10^{-11}\right) {M_\odot}/{\rm yr}\,.
\end{equation}
%
Matter in the intergalactic medium is a form of plasma that inhabits the space between galaxies, accounting for about 40-50\% of all baryons in the present Universe. Typical values for the sound speed in this medium are $(10^{-1},10^{-5}) c$~\cite{IGM1}, and a conservative estimate of the energy density is $10^{-27}\,  {\rm kg}/{\rm{m^{3}}}$~\cite{Nicastro:2007gv}. Using this and the Bondi-Hoyle rate for a supermassive ECO, we find 
\begin{equation}
\label{eq:bondi-est-IGM}
\dot{M}_{\Bondi}^{\IGM}= \left(10^{-7},10^{-19}\right) {M_\odot}/{\rm yr}\,.
\end{equation}
This estimate is consistent with the results of~\cite{IGM2}, who studied the formation of supermassive black holes using the Bondi-Hoyle rate. Finally, matter in the intracluster medium consists of superheated plasma that fills galaxy clusters, and it is composed mostly of ionized hydrogen and helium. Although this plasma accounts for most of the baryonic mass in galaxy clusters, its density is extremely low, dipping down to  $(10^{-27},10^{-31}) {\rm kg}/{\rm{m^{3}}}$~\cite{ICM1,ICM2}, which is either comparable or smaller than the dark energy density. The accretion of intracluster material by an ECO is thus negligible. 

Let us now consider the ECO accretion of dark matter. The density of dark matter depends strongly on the type of galaxy or cluster the ECO lives in. Supermassive ECOs, like supermassive black holes, are expected to live in galaxy cores, since they would sink to the center through dynamical friction. The local mass density of dark matter in our galaxy is about $10^{-21} \; {\rm{kg}}/{\rm{m}}^{3}$~\cite{Salucci:2010qr}, but near the core there is a enhancement of order $10^{4}$ (e.g.~a Navarro-Frenk-White dark halo profile \cite{NFW}). Using this, together with the average thermal velocity of cold dark matter, roughly $10^{-3} c$, one then finds
\begin{equation}
\label{eq:bondi-est}
\dot{M}_{\Bondi}^{\DM}= \left(10^{-3},10^{-7}\right) {M_\odot}/{\rm yr}\,.
\end{equation}
Reference~\cite{PPR} has shown that the Bondi estimate in Eq.~\eqref{eq:bondi-est} are very accurate for dark matter accretion.

One may also wish to consider the accretion of cosmic microwave background (CMB) radiation and dark energy into ECOs. Following the cluster plus black hole model of~\cite{PPR}, one finds that the accretion rate of CMB radiation into an $M_{0}=3000 M_\odot$ ECO is 
\begin{equation}
\dot{M}^{\CMB}_{\Bondi}= 10^{-29} \, M_\odot/{\rm yr}\,,
\end{equation}
while for dark energy, one finds
\begin{equation}
\dot{M}^{\DE}_{\Bondi}= 10^{-26}\, \left(M_{\odot}/{\rm yr}\right) \; (1+\omega)\,, 
\end{equation} 
with $p=\omega \rho$ the dark energy equation of state. Even if one scales these numbers up to a supermassive ECO, these rates are still negligible relative to dark matter accretion, and to accretion of matter in the intergalactic and interstellar medium.

\vspace{0.3cm}
\noindent
{\emph{Absorption of gravitational waves by an ECO in a binary.}} Even if ECOs are completely alone in the Universe, their mass will still increase due to the absorption of the gravitational waves they emit when in orbit around each other. Perturbation theory allows us to estimate the amount of mass-energy absorbed by a black hole (its \emph{tidal heating}) during the early inspiral of a coalescence event, and the same tools are applicable to ECOs with absorption efficiency $k$. Consider then an ECO in the early quasi-circular inspiral, where velocities are small and fields are sufficiently weak that the post-Newtonian approximation holds~\cite{Blanchet:2013haa}. The rate at which the ECO will gain mass-energy due to tidal heating is~\cite{Poisson:2005pi,Poisson:2009qj,Binnington:2009bb,Landry:2014jka}
\begin{equation}
\label{eq:mdot-tidal}
\dot{M}^{\GW}_{\tidal} = -\frac{8}{5} \, k \, \epsilon \, \eta^2 \frac{M_{0}}{m} \chi(1 + 3 \chi^2) v^{15}\, ,
\end{equation}
where $\eta = M_{0} M_{\comp}/m^{2}$ is the symmetric mass ratio, $m = M_{0}+M_{\comp}$ is the total mass, $M_{0}$ is the ECO mass and $\chi$ is its dimensionless spin, $M_{\comp}$ is the companion mass, $v$ is the orbital velocity, and $\epsilon = \pm 1$ if the spin angular momentum is aligned or anti-aligned with the orbital angular momentum. 

Unlike the case in which accretion rate is constant, here the tidal heating rate increases as the orbital velocity increases and the inspiral proceeds. Therefore, the mass gained in the inspiral cannot be estimated as explained above Eq.~\eqref{eq:Hoop}, but rather one must integrate the tidal heating rate over the relevant time scale, which in this case is the inspiral timescale. Using the Virial theorem and Kepler's third law, $v = (\pi m f)^{1/3}$, with $f$ the gravitational wave frequency, one can integrate Eq.~\eqref{eq:mdot-tidal},
\begin{align}
\label{mgainone-new}
\delta M_{\tidal}^{\GW} &= -\frac{6^{1/2} }{36288} \; \epsilon k \; \eta M_{0} \; \chi (1 + 3 \chi^2)   \, , 
\end{align}
where we have used that $df/dt = 96 \eta (\pi m f)^{11/3}/(5 \pi m^2)$, we have neglected the initial frequency since its contribution to $\delta M$ is subdominant when $f_{\rm final} \gg f_{\rm initial}$, and we have approximated the final frequency via $f_{\rm final} =  6^{-3/2}/(m \pi)$, using the innermost stable radius of a point particle on a Schwarzschild spacetime.   

We can now evaluate the above expression for a typical system. Considering then an equal-mass ECO binary with individual masses $M_{0} = M_{\comp} = 10^{6} M_{\odot}$, spins $\chi = 0.9$ anti-aligned with the orbital angular momentum, and an efficiency of $6 \%$, one finds 
\begin{align}
\label{eq:typicalGWtidal}
\delta M_{\tidal}^{\GW} &\approx 3 M_{\odot}\,,
\end{align}
during the part of the inspiral that is in the LISA band. The choice of 6\% efficiency comes from the minimum absorption required to avoid the ergoregion instability at a spin of $\chi = 0.9$; the efficiency can in fact go up to 60\% to turn off this instability for all spins~\cite{Maggio:2018ivz}.  Clearly, whether this is a mass gain or a mass loss depends on whether the spins are anti-aligned or aligned with the orbital angular momentum. If tidal heating leads to a mass loss, then the Hoop conjecture does not apply.

The arguments presented above can be generalized to non-spinning ECOs, which were considered in~\cite{Cardoso:2017cqb}. In this case, the tidal heating mass rate  is always positive 
\begin{equation}
\label{ECOs}
\dot{M}_{\tidal,\nonspin}^{\GW}=\frac{16}{5}\eta^{2}\Big(\frac{M_{0}}{m} \Big)v^{18}\,,
\end{equation}
but it is smaller than in the spinning case, as it enters at higher post-Newtonian order. The mass gained during the inspiral is now  
\begin{align}
\label{mgainone-new-no-spin}
\delta M_{\tidal,\nonspin}^{\GW} &= \frac{1}{155520} k \; \eta M_{0} \approx  10^{-1} M_{\odot} \,, 
\end{align}
where the second equality corresponds to the same system considered above.

\vspace{0.3cm}
\noindent
{\emph{The minimum radius of an ECO to prevent black hole collapse.}} We can now use the above calculations to estimate the minimum radius that an ECO must have in order to avoid collapse into a black hole due to the Hoop conjecture. For the cases that involve accretion of matter (either interstellar, intergalactic or dark), we assume the Bondi-Hoyle rate is constant to find $\delta M = \dot{M} T$, where $T$ is the amount of time between the formation of the ECO and the time it forms a binary that enters the sensitivity band of the gravitational wave detector. The timescale $T$ is somewhat uncertain, as different astrophysical process can lead to the growth of an ECO to supermassive scales, and then to the collision of two such ECOs at the center of galaxies. A reasonable time scale is roughly 1 billion years~\cite{Holley-Bockelmann:2015eua,Shibata:2011jka} (including the time for two galaxies to merge, for the supermassive objects to find each other, and for them to inspiral into the LISA band) to at most 10 billion years (the time at which quantum gravitational effects would have imprinted in the ECO production, accounting for structure formation after matter recombination). With this in mind, the amount of mass gained by accretion for the relevant processed considered here are
\begin{align}
\delta M_{\Bondi}^{\ISM} = \left(10^{0},10^{-1}\right) M_\odot \,,
& \quad
\delta M_{\Bondi}^{\IGM} = \left(10^{-3},10^{-4} \right) M_\odot \,,
\nonumber \\
\delta M_{\Bondi}^{\DM} &= \left(10^{3},10^{2}\right) M_\odot \,,
\end{align}
where to be conservative we used the non-enhanced dark matter accretion rate, since this applies only to the cusp, and ECOs will not necessarily be there for a billion years. 

Any (and all) of these processes force the ECO to have a minimum radius that is many orders of magnitude larger than a Planck length from its would-be horizon. Using Eq.~\eqref{eq:min-radius}, we find 
\begin{align}
\delta R_{\Bondi}^{\ISM} \gtrsim \left(1, 0.1\right) \; {\rm{km}}  \,,
&\quad
\delta R_{\Bondi}^{\IGM} \gtrsim \left(10^{-3},10^{-4}\right) \; {\rm{km}} \,,
\nonumber \\
\delta R_{\Bondi}^{\DM} &\gtrsim \left(10^{3},10^{2}\right) \; {\rm{km}} \,,
\end{align}
Recalling that the Planck length is $10^{-38} \; {\rm{km}}$, we see that just simple accretion processes prevent surfaces that are that close to the would-be horizon, thus making the possibility of ECOs impossible.  Similarly, ignoring matter and using only tidal heating, we find
\begin{align}
\delta R_{\tidal}^{\GW,\spinalign} &\gtrsim 1 \; {\rm{km}}  \,,
\quad
\delta R_{\tidal}^{\GW,\nonspin} \gtrsim 0.1 \;  {\rm{km}} \,,
\end{align}
over 37 orders of magnitude larger than the Planck length. Once more, tidal heating and the Hoop conjecture prevent the surface of the ECO from being a Planck length away from its would-be horizon, unless the spins are aligned with the orbital angular momentum.

From the above minimum radii one can calculate the maximum compactness of an ECO relative to that of a black hole. Using Eq.~\eqref{eq:max-comp}, we find
\begin{equation}
\frac{C}{C_{\BH}} - 1 < \frac{\delta R}{R_{\BH}} \lesssim  \frac{M_{\odot}}{M}\,.
\end{equation}
Admittedly, the right-hand side of the last inequality is a very small number for a supermassive ECO, which means that ECOs do not have to have a surface that is a Planckian distance away from the horizon to possess compactnesses that are close to that of a black hole.  

\vspace{0.3cm}
\noindent
{\emph{Discussion.}} We have shown that ECOs cannot have a surface anywhere near a Planckian distance away from their would-be horizons, if they are to avoid collapse into black holes due to accretion from the interstellar medium, intergalactic medium, dark matter accretion, or gravitational wave absorption when in a binary. This result has strong implications for the search of ECOs with tidal Love numbers~\cite{Cardoso:2017cfl,Maselli:2017cmm} or with gravitational wave echoes after the merger of two such ECOs~\cite{Abedi:2016hgu,Abedi:2017isz,Abedi:2018npz,Oshita:2018fqu}. If ECOs collapse to black holes due to accretion well before they can merge, there can be no gravitational wave signatures to begin with. This result is similar in spirit to that recently presented in~\cite{Chen:2019hfg}, where the authors considered the absorption of gravitational waves during gravitational wave echoing. Our results are in some sense stronger because they make the apply before ECOs merge.

Can one circumvent these no-go results? The easiest way to do so would be to argue that the Hoop conjecture does not apply to ECOs for some reason. This is somewhat hard to believe, since this conjecture does not rely on the particular form of the field equations. Nonetheless, one could argue that violations (though never observed) could be possible if one allows for forms of matter that do great violence to the energy conditions. If so, one would then have to explain where this strange matter came from in the first place, and how it interacts with all other astrophysical observations. Another way to bypass the no-go results is to argue that the ECO surface is not fixed, but rather it increases as its mass increases. This indeed occurs for black holes because their area is proportional to their mass squared. However, the surface of ECOs have so far been theorized to be rigid for the individual examples that exist in the literature. For instance, gravastars (ECOs with a de Sitter interior and a Schwarzschild exterior) possess a thin but rigid shell with an extremely stiff equation of state~\cite{Mazur:2001fv}. Our results, of course, do not apply to quantum ECOs that are the actual end-state of gravitational collapse in some yet-to-be-determined quantum gravity theory. 

The purpose of our analysis is not to discourage the open-minded scientific inquiry of theoretical possibilities that are not observationally excluded, but rather to point out that some of these possibilities are theoretically excluded on the basis of stability. Ultimately, rather than investigating observations that may reveal the presence of an ECO, it may be more worthwhile to cement the foundations of classical ECO theory from an action-principle stand point, thus allowing for the investigation of whether these ECOs can form dynamically in nature from gravitational collapse or not. Lacking this, ECOs remain a theoretical possibility that is currently more rooted in our imagination than in reality. 
   
\vspace{0.3cm}
\noindent
\acknowledgments {\emph{Acknowledgements}} -- We would like to thank Niayesh Afshordi, Ramy Brustein, Luis Lehner and Frans Pretorius  
for useful discussions on these subjects. 
AA and AM wish to acknowledge support by the NSFC, through the grant No. 11875113, the Shanghai Municipality, through the grant No. KBH1512299, and by Fudan University, through the grant No. JJH1512105. NY acknowledges support from NSF grant PHY-1759615 and PHY1748958, NASA grants NNX16AB98G and 80NSSC17M0041, and the hospitality of KITP where some of this work was completed. 


\end{document}